\documentclass[intlimits,twoside,a4paper]{article}

\usepackage[cp1251]{inputenc}

\usepackage[eqsecnum]{cmpj3}

\usepackage{bm}

\issue{2018}{21}{3}{33002}
\doinumber{10.5488/CMP.21.33002}

\title[Calculating spherical harmonics]%
{Calculating spherical harmonics without derivatives%
}
\author[M. Weitzman, J.K. Freericks]{M. Weitzman\refaddr{label1}, J.K. Freericks\refaddr{label2}}
\addresses{
\addr{label1} 3515 E. Rochelle Ave., Las Vegas, NV 89121, USA
\addr{label2} Department of Physics, Georgetown University, 37th and O Sts. NW, Washington, DC 20057, USA
}

\date{Received May 30, 2018}

\begin{document}

\maketitle

\begin{abstract}
The derivation of spherical harmonics is the same in nearly every quantum mechanics textbook and classroom. It 
is found to be difficult to follow, hard to understand, and challenging to reproduce by most students. In this work,
we show how one can determine spherical harmonics in a more natural way based on operators and a powerful
identity called the exponential disentangling operator identity (known in quantum optics, but little used elsewhere). This
new strategy follows naturally after one has introduced Dirac notation, computed the angular momentum algebra, and determined the action of the angular momentum raising and lowering operators on the simultaneous angular momentum eigenstates (under
$\hat L^2$ and $\hat L_z$).
\pacs 01.40.gb, 02.20.Qs, 03.65.Fd, 31.15.-p
\end{abstract}

\section{Introduction}
One of us (JKF), first met Prof. Stasyuk in the early 2000s when visiting Ukraine for a conference and then subsequently became involved in a long-standing collaboration with Andrij Shvaika, who is also at the Institute for Condensed Matter Physics in Lviv. It is always a pleasure to meet Prof. Stasyuk on subsequent trips to Lviv as his friendly smile and enthusiasm for physics is contagious. I know much of Prof. Stasyuk's career has been spent working with operator formalisms (especially Hubbard $X$-operators) and finding different ways to perform calculations or simplify complex expressions. It is within this spirit that we present this work, which provides a clear operator-based methodology for determining spherical harmonics. In spite of some of the lengthy algebraic manipulations, it is the type of derivation that can fit well into an undergraduate quantum mechanics course. For sure, it can be used in the graduate curriculum.

The motivation for this work is a book project of one of the authors (JKF) on teaching quantum mechanics without calculus~\cite{freericks_qmwc}. This project is a follow-up of a massively open on-line course called {\it Quantum Mechanics for Everyone}, which is running on edX from April 2017--April 2019~\cite{freericks_qmfe}. While for many, the idea of finding spherical harmonics without derivatives might seem sacrilegious, operator methods allow one to proceed quite far into the quantum curriculum without requiring derivatives or integrals. Indeed, most of our research work focuses on using these operator methods, so why not teach them to students from the beginning and provide a more uniform perspective on how
to perform calculations in quantum mechanics? Even quantum experts will find something new in the way we compute the spherical harmonics.

Our procedure follows a number of steps. First, we introduce the Dirac notation used to define the spherical harmonics and we analyze the required rotation operators needed to complete the calculation. Second, we derive the exponential disentangling operator identity, show its relationship to Lie algebras and Lie groups, and verify later that it works in a different representation than the one it was originally derived for. Third, we use the angular momentum algebra of raising and lowering operators to derive two equivalent (but different) formulas for the spherical harmonics. We show how one form can immediately be put into the standard format. We then analyze the second in more detail to verify that it also yields the same result. We end with a discussion of how this approach can be extended to determine the rotation matrices and provide some commentary about pedagogy.

\section{Definition of the spherical harmonics}

We work under the assumption that Dirac's bra-ket notation has already been introduced. Since this is simply a notation that emphasizes states and operators and how they inter-relate, it can be introduced early into the curriculum. We also assume that the notion of a wavefunction as the projection of a quantum state onto a coordinate-basis eigenfunction is also known.
We are working with orbital angular momentum, so the total angular momentum quantum number $l$ is a nonnegative integer $l\in \mathbb{N}$. The angular momentum operators $\hat L_x$, $\hat L_y$, and $\hat L_z$ satisfy the commutation relation
\begin{equation}
[\hat L_i,\hat L_j]=\ri\hbar\epsilon_{ijk}\hat L_k\,,
\label{eq: commutation}
\end{equation}
with $\epsilon_{ijk}$ the completely antisymmetric tensor (Levi-Civita symbol).
We also define raising and lowering operators via $\hat L_+=\hat L_x+\ri\hat L_y$ and $\hat L_-=\hat L_x-\ri\hat L_y$. These operators then satisfy the following commutation relations:
\begin{equation}
[\hat L_z,\hat L_{\pm}]=\pm\hbar\hat L_{\pm}\,, \qquad [\hat L_+,\hat L_-]=2\hbar\hat L_z.
\label{eq: commutation2}
\end{equation}

We then start with the angular momentum eigenstates $|l,m\rangle$ given by the eigenvalue-eigenvector relations
\begin{equation}
\hat L^2|l,m\rangle=(\hat L_x^2+\hat L_y^2+\hat L_z^2)|l,m\rangle=\hbar^2 l(l+1)|l,m\rangle
\label{eq: l_squared}
\end{equation}
and
\begin{equation}
\hat L_z|l,m\rangle=\hbar m |l,m\rangle.
\label{eq: l_z}
\end{equation}
Next, we need to determine the angular coordinate-basis state $|\theta,\phi\rangle$, where $\theta$ is the angle of the spherical coordinate with respect to the $z$-axis and $\phi$ is the azimuthal angle rotating around the $z$-axis. Using the angular momentum operators to represent the rotations about the $y$ and $z$-axes, we can express this state as
\begin{equation}
|\theta,\phi\rangle=\re^{-\ri\phi\frac{\hat L_z}{\hbar}}\re^{-\ri \theta\frac{\hat L_y}{\hbar}}|\theta=0,\phi=0\rangle,
\label{eq: angle_ket}
\end{equation}
which expresses the general state as the rotation of the state that starts oriented toward the north pole along the $z$-axis --- we rotate along the $y$-axis by the angle $\theta$ and along the $z$-axis by the angle $\phi$ to reach the $|\theta,\phi\rangle$ state. We compute the wavefunction (or the spherical harmonic), by simply taking the hermitian conjugate of the coordinate ket state (to make it a bra) and overlapping it with the angular-momentum eigenstate. This gives
\begin{equation}
Y_l^m(\theta,\phi)=\langle\theta,\phi|l,m\rangle=\langle\theta=0,\phi=0|\re^{\ri\theta\frac{\hat L_y}{\hbar}}\re^{\ri\phi\frac{\hat L_z}{\hbar}}|l,m\rangle,
\label{eq: spherical_harmonic_def}
\end{equation}
where we use the standard notation $Y_l^m(\theta,\phi)$ for the spherical harmonic wavefunction.
Since the state $|l,m\rangle$ is an eigenstate of $\hat L_z$, the first operator can be immediately evaluated as $\exp(\ri m\phi)$. We introduce a complete set of states in between the coordinate bra on the left and the remaining rotation operator. This complete set of states includes only states with the same $l$ eigenvalue, because the $\hat L^2$ operator commutes with the rotation operator (implying that the overlap with any angular-momentum eigenstate from a different multiplet vanishes). This then yields
\begin{equation}
Y_l^m(\theta,\phi)=\re^{\ri m\phi}\sum_{m'=-l}^l\langle\theta=0,\phi=0|l,m'\rangle\langle l,m'|\re^{\ri\theta\frac{\hat L_y}{\hbar}}|l,m\rangle,
\label{eq: spherical_harmonic2}
\end{equation}
which experts will recognize as involving Wigner's rotation matrix in the angular momentum multiplet corresponding to total angular momentum $l$. 

We next note that the ``north pole'' state $|\theta=0,\phi=0\rangle$ must have zero overlap with any $|l,m\rangle$ state that has $m\ne 0$. This is because the matrix element 
$\langle \theta=0,\phi=0|l,m\rangle$ 
is equal to $\langle\theta=0,\phi=0|\re^{\ri\phi'\hat L_z/\hbar}|l,m\rangle$, since the north-pole state is unchanged by any rotation about the $z$-axis. This implies that we must have $m=0$, because that is the only eigenstate in the angular momentum multiplet that also is unchanged by a rotation along the $z$-axis.

We pause here for a moment to give an important comment. It is often claimed that orbital angular momentum must involve an integer eigenvalue for the total angular momentum because this is the only way to have the wavefunction be continuous as we wind the angle $\phi$ by $2\piup$. But such a condition is not required, since it is only the probability density that must be single-valued, not the wavefunction~\cite{green,ballentine}. Nevertheless, the above argument, that we  can only find a wavefunction by expanding in terms of the $m'=0$ row of the rotation matrix, naturally implies that $l$ must be a nonnegative integer. Otherwise, there is no $m'=0$ component!

This observation reduces our computation of the spherical harmonics to the final identity
\begin{equation}
Y_l^m(\theta,\phi)=\re^{\ri m\phi}\langle\theta=0,\phi=0|l,m'=0\rangle\langle l,m'=0|\re^{\ri\theta\frac{\hat L_y}{\hbar}}|l,m\rangle.
\label{eq: spherical_harmonic_final}
\end{equation}
Note that the factor $\langle\theta=0,\phi=0|l,m'=0\rangle$ is just an overall normalization constant. The difficulty lies in calculating the matrix element of the rotation operator. One can see the challenge by expressing $\hat L_y=(\hat L_+-\hat L_-)/2\ri$ and recognizing that one cannot simply express this as a product of exponentials of the raising and lowering operators multiplied by a correction term because the correction term from the Baker-Campbell-Hausdorff expression will never terminate. Instead, we will employ the exponential disentangling operator identity to determine a simple operator form that can be evaluated. As you will see below, this powerful identity should become part of the standard quantum mechanics curriculum.

\section{Exponential disentangling operator identity}

In a seminal paper by Arecchi, Courtens, Gilmore, and Thomas from 1972~\cite{arecchi}, many different relationships between harmonic oscillator coherent states and angular momentum coherent states are discussed. In particular, the exponential disentangling operator identity is fully derived in their appendix~A. 

Before we get into the details of the derivation, we want to first briefly describe what a Lie algebra and a Lie group are. We will not give the most general mathematical definition, but rather describe a working definition that is appropriate for quantum mechanics. A Lie algebra consists of a set of operators that are closed under commutation relations. An example is the angular momentum operators. One key property is that the Lie algebra satisfies the Jacobi identity, which we  know every commutator satisfies. We form the Lie group from exponentiating scalars multiplied by the Lie algebra operators in arbitrary linear combinations. The product of any two Lie group elements can then be expressed as the exponential of a linear combination of the Lie algebra operators due to the Baker-Campbell-Hausdorf theorem and the fact that the Lie algebra closes under commutation. Hence, the Lie group is a mathematical group that has multiplication, a unique inverse, etc. For physics, we often focus on finite-dimensional representations of Lie groups and Lie algebras (Ado's theorem~\cite{ado} guarantees that such representations exist with finite-dimensional matrices). In this case, a faithful representation is a representation that is injective (or one-to-one) so that every element of the Lie group is mapped to a different element of the representation. Faithful representations also have a trivial null space. The theory of representations then tells us that any identity written in terms of exponentials of the Lie algebra operators that one can derive in a faithful representation will hold both at the general operator level and for all other faithful representations (see for example \cite{fernandez_castro}). This follows, essentially, from the definition of a faithful representation, due to the one-to-one nature of the mapping. We use this important fact in the derivation of the exponential disentangling operator identity.

We employ the simplest representation of angular momentum, namely the results for $l=1/2$ with $2\times 2$ matrices. The representation of the angular momentum matrices is
\begin{equation}
\hat L_x=\frac{\hbar}{2}\left ( \begin{array}{c c} 0&1\\1&0\end{array}\right )=\frac{\hbar}{2}\sigma_x\,, \qquad
\hat L_y=\frac{\hbar}{2}\left ( \begin{array}{c c} 0&-\ri\\\ri&0\end{array}\right )=\frac{\hbar}{2}\sigma_y\,, \qquad
\hat L_z=\frac{\hbar}{2}\left ( \begin{array}{c c} 1&0\\0&-1\end{array}\right )=\frac{\hbar}{2}\sigma_z\,,
\label{eq: pauli}
\end{equation}
where we continue to use $\hat L$ to represent the angular momentum, even though this case is for spin one-half. We also find that
\begin{equation}
\hat L^2=\hbar^2\left ( \begin{array}{c c} \frac{3}{4}&0 \vspace{1mm}\\ 
0&\frac{3}{4}\end{array}\right ), \qquad
\hat L_+=\frac{\hbar}{2}\left ( \begin{array}{c c} 0&2\\0&0\end{array}\right )=\frac{\hbar}{2}\sigma_+\,, \qquad
\hat L_-=\frac{\hbar}{2}\left ( \begin{array}{c c} 0&0\\2&0\end{array}\right )=\frac{\hbar}{2}\sigma_-\,.
\label{eq: pauli2}
\end{equation}
The Pauli matrices $\sigma_i$ satisfy $\sigma_i^2=\mathbb{I}$, so we can immediately derive the fundamental exponential identity
\begin{equation}
\exp\left ( \ri\alpha\vec{n}\cdot\vec{\sigma}\right )=\cos\alpha\,\mathbb{I}+\ri\sin\alpha\, \vec{n}\cdot\vec{\sigma},
\label{eq: exp_pauli}
\end{equation}
where $\vec{n}$ is a unit vector and $\vec{n}\cdot\vec{\sigma}=\sum_{i=1}^3 n_i\sigma_i$. The derivation proceeds simply by expanding the exponential in a power series and using the fact that
the Pauli matrices square to the identity. Note that this identity works in two ways: first, it shows us how to go from the exponential form of a Lie group element to its $2\times 2$ matrix form and second, it shows us how to convert any $2\times 2$ matrix into an exponential Lie group form.

We are now ready to derive the exponential disentangling operator identity. We have for spin one-half
\begin{equation}
\re^{\ri\theta\frac{\hat L_y}{\hbar}}=\re^{\ri\frac{\theta}{2}\sigma_y}=\cos\left (\frac{\theta}{2}\right ) \,\mathbb{I}+\ri\sin\left (\frac{\theta}{2}\right )\sigma_y=\left (\begin{array}{c c}\cos\left (\frac{\theta}{2}\right )&\sin\left (\frac{\theta}{2}\right) \vspace{1mm}\\
-\sin\left (\frac{\theta}{2}\right )&\cos\left (\frac{\theta}{2}\right )\end{array}\right ).
\label{eq: exp_ly}
\end{equation}
We wish to factorize this in the following form
\begin{equation}
\re^{\ri\theta\frac{\hat L_y}{\hbar}}=\left ( \begin{array}{cc} 1&0\\a&1\end{array}\right )
\left ( \begin{array}{cc} b&0\\0&c\end{array}\right )
\left ( \begin{array}{cc} 1&d\\0&1\end{array}\right )=
\left ( \begin{array}{cc} b&bd\\ab&abd+c\end{array}\right ),
\label{eq: disent1}
\end{equation}
which immediately tells us that $a=-\tan(\theta/2)$, $b=\cos(\theta/2)$, $c=\sec(\theta/2)$, and $d=\tan(\theta/2)$. This allows us to re-express the factorization in terms of exponentials of matrices via
\begin{equation}
\re^{\ri\theta\frac{\hat L_y}{\hbar}}=\re^{-\frac{1}{2}\tan\left (\frac{\theta}{2}\right )\sigma_-}
\re^{\ln\left [ \cos\left (\frac{\theta}{2}\right )\right ]\sigma_z}
\re^{\frac{1}{2}\tan\left (\frac{\theta}{2}\right )\sigma_+}
\label{eq: disent2}
\end{equation}
and we do not need to worry about the absolute value of the logarithm because $0\leqslant\theta\leqslant\piup$.
This latter form follows because $\sigma_{\pm}$ are nilpotent matrices, which satisfy $\sigma_{\pm}^2=0$ or
\begin{equation}
\re^{\alpha \sigma_+}=\left (\begin{array}{c c} 1 & 2\alpha\\0&1\end{array}\right ) \quad \text{and} \quad \re^{\alpha\sigma_-}=\left (\begin{array}{c c} 1&0\\2\alpha&1\end{array}\right ).
\label{eq: exp_pm}
\end{equation}
We now re-express this identity in terms of the Lie algebra, and claim it holds for all representations. Namely, we have
\begin{equation}
\re^{\ri\theta\frac{\hat L_y}{\hbar}}=\re^{-\tan\left (\frac{\theta}{2}\right )\frac{\hat L_-}{\hbar}}
\re^{\ln\left [ \cos^2\left (\frac{\theta}{2}\right )\right ]\frac{\hat L_z}{\hbar}}
\re^{\tan\left (\frac{\theta}{2}\right )\frac{\hat L_+}{\hbar}}
\label{eq: disent3}
\end{equation}
as the first form for the exponential disentangling operator identity. Note that the above strategy works with all Lie algebras/groups and can be employed to determine more complicated identities than the above one. It seems that much of the confusion about these identities arises from the fact that they were derived for a particular representation and then the claim is that they hold for all representations. While this must be true simply from the definition of a faithful representation~\cite{fernandez_castro}, the result can appear ``magical''. 

The second identity simply reverses the order of the raising and lowering operator factors. Hence, we factorize
\begin{equation}
\re^{\ri\theta\frac{\hat L_y}{\hbar}}=\left ( \begin{array}{cc} 1&a'\\0&1\end{array}\right )
\left ( \begin{array}{cc} b'&0\\0&c'\end{array}\right )
\left ( \begin{array}{cc} 1&0\\d'&1\end{array}\right )=
\left ( \begin{array}{cc} b'+a'c'd'&a'c'\\c'd'&c'\end{array}\right ),
\label{eq: disent4}
\end{equation}
to find $a'=\tan(\theta/2)$, $b'=\sec(\theta/2)$, $c'=\cos(\theta/2)$, and $d'=-\tan(\theta/2)$. This then yields our second exponential disentangling operator identity:
\begin{equation}
\re^{\ri\theta\frac{\hat L_y}{\hbar}}=\re^{\tan\left (\frac{\theta}{2}\right )\frac{\hat L_+}{\hbar}}
\re^{-\ln\left [ \cos^2\left (\frac{\theta}{2}\right )\right ]\frac{\hat L_z}{\hbar}}
\re^{-\tan\left (\frac{\theta}{2}\right )\frac{\hat L_-}{\hbar}}.
\label{eq: disent5}
\end{equation}
Although these two identities are simple to derive, they end up being quite powerful to use in a number of different applications. They should be better known outside the realm of quantum optics. Perhaps this will occur if others use them to teach about spherical harmonics. Pressing forward, we use them to complete our spherical harmonics derivation next.

\section{Derivation of the spherical harmonics from the angular momentum algebra}

We now have all of the elements needed to determine the spherical harmonics. Because the exponential disentangling operator identity has two forms, we have two ways we can proceed. We begin by working with the identity that has the raising operator to the right and the lowering to the left. Then, we find
\begin{equation}
Y^m_l(\theta,\phi)=\re^{\ri m\phi}\langle \theta=0,\phi=0 | l,m'=0\rangle\langle l,m'=0|\re^{\ri\theta\frac{\hat{L}_y}{\hbar}}|l,m\rangle
\end{equation}
becomes
\begin{equation}
Y^m_l(\theta,\phi)=\re^{\ri m\phi}\langle \theta=0,\phi=0 | l,m'=0\rangle\langle l,m'=0|
\re^{-\tan\left (\frac{\theta}{2}\right )\frac{\hat L_-}{\hbar}}
\re^{\ln\left [ \cos^2\left (\frac{\theta}{2}\right )\right ]\frac{\hat L_z}{\hbar}}
\re^{\tan\left (\frac{\theta}{2}\right )\frac{\hat L_+}{\hbar}}|l,m\rangle.
\end{equation}

\begin{figure}[!b]
\centerline{\includegraphics[width=0.6\textwidth]{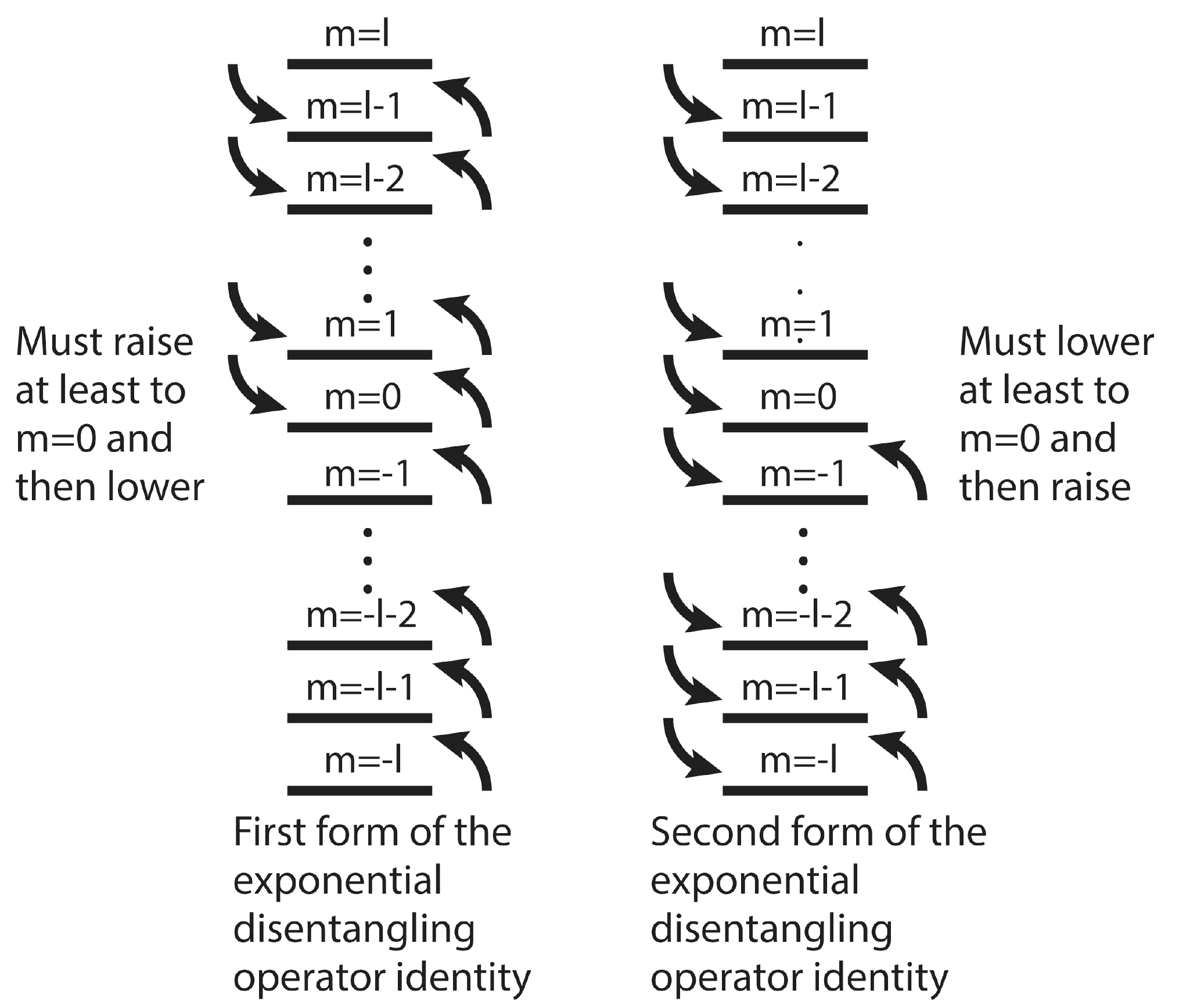}}
\caption{Schematic figure indicating the raising and lowering operators for the two forms of the exponential disentangling operator identity in equations~(\ref{eq: disent3}) and (\ref{eq: disent5}) (left and right, respectively). Since we must end at $m=0$, the first form requires us to raise states (arrows moving up on the right) that start with $m<0$ to at least $m=0$ (if raised higher, they must be lowered down to $m=0$ indicated by the arrows going down on the left). Of course, there is no constraint for states that start with $m\geqslant 0$. Similarly for the second form, we must lower an $m>0$ state to at least $m=0$ (arrows moving down on the left) with no constraint for the states that start with $m\leqslant 0$. We only illustrate the ladder operators action with the arrows --- we also need to attach the $\hat L_z$ contributions along the way. The exponentials produce polynomials in the raising and lowering operators that involve all possible number of steps up and down (or \textit{vice versa})  that start from the initial $m$ value and end at $m=0$. \label{fig: schematic_mzp}}
\end{figure}

We can immediately analyze the strategy for completing the calculation, which is illustrated schematically in figure~\ref{fig: schematic_mzp}. Because the highest-weight state cannot be raised --- i.e., $\hat L_+|l,m=l\rangle=0$ --- the exponential of the raising operator acting on the state on the right will involve exactly $l-m+1$ terms in a polynomial in $\hat L_+$. A similar result is true for the exponential of the  $\hat L_-$ term. Furthermore, the Lie algebra for the angular momentum operators immediately tells us what each term is from repeatedly using the raising and lowering operator matrix elements
\begin{equation}
\hat L_+|l,m\rangle=\hbar\sqrt{(l-m)(l+m+1)}|l,m+1\rangle \quad \textrm{and} \quad
\hat L_-|l,m\rangle=\hbar\sqrt{(l+m)(l-m+1)}|l,m-1\rangle,
\label{eq: raise_lower}
\end{equation}
which we assume have already been derived prior to working out the spherical harmonics wavefunctions. Hence, we simply employ equation~(\ref{eq: raise_lower}) multiple times to determine each term in the series. For example, we immediately find
\begin{equation}
\re^{\tan\left (\frac{\theta}{2}\right )\frac{\hat L_+}{\hbar}}|l,m\rangle=
\sum_{n=0}^{l-m}\frac{\left ( \tan\frac{\theta}{2}\right )^n}{n!}\left [\prod_{r=1}^n\sqrt{(l-m-r+1)(l+m+r)}\right ]|l,m+n\rangle.
\end{equation}
Operating the middle term next yields
\begin{equation}
\re^{\ln\left [ \cos^2\left (\frac{\theta}{2}\right )\right ]\frac{\hat L_z}{\hbar}}\re^{\tan\left (\frac{\theta}{2}\right )\frac{\hat L_+}{\hbar}}|l,m\rangle=
\sum_{n=0}^{l-m}\frac{\left ( \tan\frac{\theta}{2}\right )^n}{n!}\left [\prod_{r=1}^n\sqrt{(l-m-r+1)(l+m+r)}\right ] \left (\cos\frac{\theta}{2}\right )^{2m+2n}|l,m+n\rangle.
\end{equation}
The last step is to operate the final factor with the lowering operator on this state. Because we take the overlap with $\langle l,m'=0|$, we only need to include the term that applies the 
$\hat L_-$ operator $m+n$ times. We also need to have $m+n\geqslant 0$, otherwise, we cannot lower to the state with $m'=0$. This then gives us
\begin{align}
\langle l,m'=0|\re^{\ri\theta\frac{\hat L_y}{\hbar}}|l,m\rangle&=
\sum_{n=\max(0,-m)}^{l-m}
\frac{\left (- \tan\frac{\theta}{2}\right )^{m+n}}{(m+n)!}
\frac{\left ( \tan\frac{\theta}{2}\right )^n}{n!}
\left [\prod_{r=1}^n\sqrt{(l-m-r+1)(l+m+r)}\right ]\nonumber\\
&\times\left (\cos\frac{\theta}{2}\right )^{2m+2n}
\left [\prod_{s=1}^{m+n}\sqrt{(l+m+n-s+1)(l-m-n+s)}\right ],
\end{align}
where we used the fact that $\langle l,m=0|l,m=0\rangle=1$. This final result can be simplified.
Start with the trigonometric factors. Using the half-angle identities $\sin\theta=2\sin(\theta/2)\cos(\theta/2)$ and $1-\cos\theta=2\sin^2(\theta/2)$ then gives
\begin{equation}
\left (- \tan\frac{\theta}{2}\right )^{m+n}\left ( \tan\frac{\theta}{2}\right )^n\left (\cos\frac{\theta}{2}\right )^{2m+2n}=\left (-\frac{1}{2}\right )^{m+n}\left (\sin\theta\right )^m\left ( 1-\cos\theta\right )^n.
\end{equation}
The two product terms become
\begin{equation}
\sqrt{\frac{(l-m)!(l+m+n)!(l+m+n)!l!}{(l-m-n)!(l+m)!l!(l-m-n)!}}=\frac{(l+m+n)!}{(l-m-n)!}\sqrt{\frac{(l-m)!}{(l+m)!}}.
\end{equation}
Using these two results, then yields
\begin{equation}
\langle l,m'=0|\re^{\ri\theta\frac{\hat L_y}{\hbar}}|l,m\rangle=\sum_{n=\max(0,-m)}^{l-m}
\left (-\frac{1}{2}\right )^{m+n}\left (\sin\theta\right )^m\left ( 1-\cos\theta\right )^n
\frac{(l+m+n)!}{(m+n)!n!(l-m-n)!}\sqrt{\frac{(l-m)!}{(l+m)!}}.
\label{eq: mzp_final}
\end{equation}
This is multiplied by $\exp(\ri m \phi)\langle \theta=0,\phi=0|l,m'=0\rangle$ to finally yield the spherical harmonic.

However, we do need to put this result into a standard form, which we can directly do for
$m\geqslant 0$. In this case, we employ the definition of the associated Legendre function via~\cite{legendre1}
\begin{equation}
P_l^m(\cos\theta)=\frac{1}{2^mm!}\frac{(l+m)!}{(l-m)!}(\sin\theta)^m\sum_{n=0}^{l-m}(-1)^n\frac{\left ( \begin{array}{c} 
l-m\\n
\end{array}\right )\left ( \begin{array}{c}
l+m+n\\n
\end{array}\right )}{\left (\begin{array}{c}
m+n\\n
\end{array}\right )}\left (\frac{1-\cos\theta}{2}\right )^n,
\end{equation}
where the combinatorial symbol satisfies
\begin{equation}
\left (
\begin{array}{c}
m\\
n
\end{array}\right )=\frac{m!}{(m-n)!n!}.
\end{equation}
After evaluating the combinatorial factors, one can immediately see that
\begin{equation}
P_l^m(\cos\theta)=\frac{1}{2^m}(\sin\theta)^m\sum_{n=0}^{l-m}(-1)^n\frac{(l+m+n)!}{(l-m-n)!(m+n)!n!}\left (\frac{1-\cos\theta}{2}\right )^n
\end{equation}
and hence
\begin{equation}
Y_l^m(\theta,\phi)=\langle \theta=0,\phi=0|l,m'=0\rangle(-1)^m\sqrt{\frac{(l-m)!}{(l+m)!}}P_l^m(\cos\theta)\re^{\ri m\phi},
\end{equation}
for $m\geqslant 0$. If we proceed in the standard fashion, the case $m<0$ follows from the general relation $Y_l^{-|m|}(\theta,\phi)=(-1)^{|m|}[Y_l^{|m|}(\theta,\phi)]^*$. The term $\langle \theta=0,\phi=0|l,m'=0\rangle$ is a normalization factor, which can be shown to be equal to $\sqrt{(2l+1)/4\piup}$ by normalizing the wavefunction. 

If we instead determine $Y_l^m$ directly from the formula derived above for $m<0$ in equation~(\ref{eq: mzp_final}), we see that it appears to be a new formula, as it does not seem to have any simple relationship with the conventional formula. In particular, we examine the case $l=1$ and $m=-1$. Here, we have 
\begin{equation}
Y_1^{-1}(\theta,\phi)=\sqrt{\frac{3}{4\piup}}\re^{-\ri \phi}\left [
\frac{1-\cos\theta}{\sin\theta}\sqrt{2}-\frac{1}{2}\frac{(1-\cos\theta)^2}{\sin\theta}\sqrt{2}\right ]=\sqrt{\frac{3}{8\piup}}\re^{-\ri\phi}\frac{1-\cos^2\theta}{\sin\theta}=
\sqrt{\frac{3}{8\piup}}\re^{-\ri\phi}\sin\theta,
\end{equation}
which is the correct result, but requires a nontrivial cancellation of $\sin\theta$ in the numerator and denominator. This brings to bear the question of whether the result for $m<0$ is
a new result, or the one that reduces to the known result after using the special properties of the associated Legendre functions. We discuss this in detail below and show that it is not a new identity.

But first, we derive a similar result for the other form of the exponential disentangling operator identity. 
Here, the raising operator appears on the left and the lowering operator on the right. Otherwise,
the procedure is precisely the same. We simply report the final result analogous to
equation~(\ref{eq: mzp_final}):
\begin{align}
\langle l,m'=0|\re^{\ri\theta\frac{\hat L_y}{\hbar}}|l,m\rangle&=\sum_{n=\max(0,m)}^{l+m}
(-1)^{m}\left (-\frac{1}{2}\right )^{n-m}\left (\sin\theta\right )^{-m}\left ( 1-\cos\theta\right )^n\nonumber\\
&\times\frac{(l-m+n)!}{(n-m)!n!(l+m-n)!}
\sqrt{\frac{(l+m)!}{(l-m)!}}.
\label{eq: pzm_final}
\end{align}
So the result for the second form of the exponential disentangling operator identity is the same as the first form with $m\to-m$ and an additional phase factor of $(-1)^{m}$. Note how, in this case, the results for $m\leqslant 0$ are the conventional results, while those for $m>0$
appear to be new formulas. Note further, that the symmetry of $m\to -m$ and the extra factor of $(-1)^m$, are exactly what is needed to verify the aforementioned identity, that $Y_l^m(\theta,\phi)=(-1)^mY_l^{-m*}(\theta,\phi)$, which is usually derived by taking a complex conjugate of the differential equation. Of course, the final evaluation of the results derived from the two different forms of the exponential disentangling operator identity must be equal to each other for the corresponding $m$ values.

We now return to equation~\eqref{eq: mzp_final} for the case $m<0$.  That this is equivalent to the conventional spherical harmonics formula is essentially shown in appendix~D of \cite{arecchi} where their equations (D5) and (D6) are shown to be equivalent.  But the form that those equations take, being the full rotation matrices, is not the most direct way to show that the new forms of the identities actually reduce to the conventional ones. We know that this equivalence is highly nontrivial, given the example we worked out for $l=1$ above. Hence, we do not recommend going through the details of the following derivation with students, as it is likely to be too much algebra for them to be able to follow all of the steps. But for completeness, we do include all of those details here.

It turns out that the strategy we use is to start with the first form of the exponential disentangling operator identity for $m<0$, apply the methodology described in \cite{arecchi} to arrive at an intermediate form, similar to their equation (D6), and then apply the methodology in reverse, to show that the intermediate result
also agrees with the first form of the identity with $m>0$ [up to an additional factor of $(-1)^m$], which then shows us the equivalence of the negative $m$ and positive $m$ results. We assume $m<0$, so $m=-|m|$. We begin with equation~\eqref{eq: mzp_final} and rewrite the  trigonometric functions in terms of their half-angle identities, to yield
\begin{align}
\langle l,m'=0|\re^{\ri\theta\frac{\hat L_y}{\hbar}}|l,-|m|\rangle&=\sum_{n=|m|}^{l+|m|}
\left (-1\right )^{-|m|+n}\left (\sin\frac{\theta}{2}\right )^{2n-|m|}\left ( \cos\frac{\theta}{2}\right )^{-|m|}\nonumber
\end{align}
\begin{align}
\times\frac{(l-|m|+n)!}{(-|m|+n)!n!(l+|m|-n)!}\sqrt{\frac{(l+|m|)!}{(l-|m|)!}}.
\end{align}
We then shift $n\to n+|m|$ to find
\begin{equation}
\langle l,m'=0|\re^{\ri\theta\frac{\hat L_y}{\hbar}}|l,-|m|\rangle=\sum_{n=0}^{l}
\left (-1\right )^{n}\left (\sin\frac{\theta}{2}\right )^{2n+|m|}\left ( \cos\frac{\theta}{2}\right )^{-|m|}
\frac{(l+n)!}{n!(n+|m|)!(l-n)!}\sqrt{\frac{(l+|m|)!}{(l-|m|)!}}.
\end{equation}
The next step uses the van der Monde identity, given by
\begin{equation}
\sum_{k=0}^{\min(r,m)} \left (\begin{array}{c} m\\k\end{array}\right )
\left (\begin{array}{c}n\\r-k\end{array}\right )=\left (\begin{array}{c}n+m\\r\end{array}\right ),
\end{equation}
where $m$, $n$, $r$, and $k$ are all nonnegative integers, and we will assume that $n>r$. The first form we use the  identity in, after some rearranging of terms, is
\begin{equation}
\frac{(l+n)!}{n!(l-|m|)!}=\sum_{k=0}^{\min(n,l-|m|)}\frac{l!(n+|m|)!}{k!(l-|m|-k)!(n-k)!(|m|+k)!}.
\end{equation}
Substituting this in, then gives
\begin{align}
\langle l,m'=0|\re^{\ri\theta\frac{\hat L_y}{\hbar}}|l,-|m|\rangle&=\sum_{n=0}^{l}\sum_{k=0}^{\min(n,l-|m|)}
\left (-1\right )^{n}\left (\sin\frac{\theta}{2}\right )^{2n+|m|}\left ( \cos\frac{\theta}{2}\right )^{-|m|}\nonumber\\
&\times
\frac{l!\sqrt{(l+|m|)!(l-|m|)!}}{k!(l-|m|-k)!(n-k)!(|m|+k)!(l-n)!}.
\end{align}
Switching the order of the $k$ and $n$ sums yields
\begin{align}
\langle l,m'=0|\re^{\ri\theta\frac{\hat L_y}{\hbar}}|l,-|m|\rangle&=\sum_{k=0}^{l-|m|} \sum_{n=k}^{l}
\left (-1\right )^{n}\left (\sin\frac{\theta}{2}\right )^{2n+|m|}\left ( \cos\frac{\theta}{2}\right )^{-|m|}\nonumber\\
&\times
\frac{l!\sqrt{(l+|m|)!(l-|m|)!}}{k!(l-|m|-k)!(n-k)!(|m|+k)!(l-n)!}.
\end{align}
Now, we shift $n\to n+k$ to find
\begin{align}
\langle l,m'=0|\re^{\ri\theta\frac{\hat L_y}{\hbar}}|l,-|m|\rangle&=\sum_{k=0}^{l-|m|} \left (-1\right )^{k}\left (\sin\frac{\theta}{2}\right )^{2k+|m|}\left ( \cos\frac{\theta}{2}\right )^{-|m|}
\frac{l!\sqrt{(l+|m|)!(l-|m|)!}}{k!(l-|m|-k)!(|m|+k)!}\nonumber\\
&\times
\sum_{n=0}^{l-k}
\left (-1\right )^{n}\left (\sin\frac{\theta}{2}\right )^{2n}
\frac{1}{n!(l-k-n)!}.
\end{align}
The sum over $n$ can now be done, yielding a $[1-\sin^2(\theta/2)]=\cos^2(\theta/2)$ raised to the $l-k$ power and a factorial term, as follows
\begin{equation}
\langle l,m'=0|\re^{\ri\theta\frac{\hat L_y}{\hbar}}|l,-|m|\rangle=\sum_{k=0}^{l-|m|} \left (-1\right )^{k}\left (\sin\frac{\theta}{2}\right )^{2k+|m|}\left ( \cos\frac{\theta}{2}\right )^{2l-2k-|m|}
\frac{l!\sqrt{(l+|m|)!(l-|m|)!}}{k!(l-|m|-k)!(|m|+k)!(l-k)!}.
\end{equation}

This is the middle point of our derivation and corresponds to the formula (D6) from \cite{arecchi}. The rest of the derivation proceeds in a similar fashion backwards, reversing all of the steps, but making changes along the way to produce the $m>0$ result.

Okay, here we go. We start by factoring out a term $[\cos^2(\theta/2)]^{l-|m|-k}=[1-\sin^2(\theta/2)]^{l-|m|-k}$ and then expanding the term in the binomial expansion to give
\begin{align}
\langle l,m'=0|\re^{\ri\theta\frac{\hat L_y}{\hbar}}|l,-|m|\rangle&=\sum_{k=0}^{l-|m|} \left (-1\right )^{k}\left (\sin\frac{\theta}{2}\right )^{2k+|m|}\left ( \cos\frac{\theta}{2}\right )^{|m|}
\frac{l!\sqrt{(l+|m|)!(l-|m|)!}}{k!(l-|m|-k)!(|m|+k)!(l-k)!}\nonumber\\
&\times\sum_{n=0}^{l-|m|-k}(-1)^n\left (\sin^2\frac{\theta}{2}\right )^{n}\frac{(l-|m|-k)!}{n!(l-|m|-k-n)!}.
\end{align}
Now, we shift $n\to n-k$:
\begin{align}
\langle l,m'=0|\re^{\ri\theta\frac{\hat L_y}{\hbar}}|l,-|m|\rangle&=\sum_{k=0}^{l-|m|} \left (\sin\frac{\theta}{2}\right )^{2k+|m|}\left ( \cos\frac{\theta}{2}\right )^{|m|}
\frac{l!\sqrt{(l+|m|)!(l-|m|)!}}{k!(|m|+k)!(l-k)!}\nonumber\\
&\times\sum_{n=k}^{l-|m|}(-1)^{n}\left (\sin^2\frac{\theta}{2}\right )^{n-k}\frac{1}{(n-k)!(l-|m|-n)!}.
\end{align}
Now, we switch the order of the $k$ and $n$ sums to give
\begin{align}
\langle l,m'=0|\re^{\ri\theta\frac{\hat L_y}{\hbar}}|l,-|m|\rangle&=\sum_{n=0}^{l-|m|}\sum_{k=0}^{n} (-1)^{n}\left (\sin\frac{\theta}{2}\right )^{2n+|m|}\left ( \cos\frac{\theta}{2}\right )^{|m|}\nonumber\\
&\times
\frac{l!\sqrt{(l+|m|)!(l-|m|)!}}{k!(|m|+k)!(l-k)!(n-k)!(l-|m|-n)!}.
\end{align}
We next shift $k\to k-|m|$:
\begin{align}
\langle l,m'=0|\re^{\ri\theta\frac{\hat L_y}{\hbar}}|l,-|m|\rangle&=\sum_{n=0}^{l-|m|}\sum_{k=|m|}^{n+|m|} (-1)^{n}\left (\sin\frac{\theta}{2}\right )^{2n+|m|}\left ( \cos\frac{\theta}{2}\right )^{|m|}\nonumber\\
&\times
\frac{l!\sqrt{(l+|m|)!(l-|m|)!}}{(k-|m|)!k!(l-k+|m|)!(n-k+|m|)!(l-|m|-n)!}.
\end{align}
As before, we now introduce the van der Monde identity in the form
\begin{equation}
\frac{(l+n+|m|)!}{n!(l+|m|)!}=\sum_{k=|m|}^{n+|m|}\frac{l!(n+|m|)!}{(k-|m|)!k!(l-k+|m|)!(n-k+|m|)!}
\end{equation}
and find
\begin{equation}
\langle l,m'=0|\re^{\ri\theta\frac{\hat L_y}{\hbar}}|l,-|m|\rangle=\sum_{n=0}^{l-|m|} (-1)^{n}\left (\sin\frac{\theta}{2}\right )^{2n+|m|}\left ( \cos\frac{\theta}{2}\right )^{|m|}
\frac{(l+n+|m|)!}{(n+|m|)!n!(l-|m|-n)!}\sqrt{\frac{(l-|m|)!}{(l+|m|)!}}.
\end{equation}
Finally, using the half-angle formulas, again gives our final result
\begin{align}
\langle l,m'=0|\re^{\ri\theta\frac{\hat L_y}{\hbar}}|l,-|m|\rangle&=(-1)^{|m|}\sum_{n=0}^{l-|m|} \left ( -\frac{1}{2}\right )^{|m|+n}\left (\sin\theta\right )^{|m|}\left (1- \cos\theta\right )^{n}\nonumber\\
&\times\frac{(l+n+|m|)!}{(n+|m|)!n!(l-|m|-n)!}
\sqrt{\frac{(l-|m|)!}{(l+|m|)!}}.
\end{align}
Hence, we have shown that the result for $m=-|m|$ is the same as the result for $m=|m|$ 
with an additional factor of $(-1)^{|m|}$. This is exactly what is needed to show $Y_l^m(\theta,\phi)=(-1)^mY_l^{-m*}(\theta,\phi)$.

This derivation was tortuously long, but we could not find any shortcut to do it faster.
It is possible that some of the identities for the Jacobi polynomials could be employed to show this 
directly, but we have not been able to sort that out here.

\section{Pedagogical aspects}

Historically, there are five ways that spherical harmonics can be derived. The first way, and the oldest, is from the works of Laplace~\cite{laplace} and Legendre~\cite{legendre} who determined the spherical harmonics by solving the differential equation. The only change, when one goes to quantum mechanics, is that we now also have the normalization condition.  The second way is via harmonic polynomials, a method first developed by Kramers~\cite{kramers} and most recently popularized by Weinberg~\cite{weinberg}. The third, is to use the highest weight state $|l,l\rangle$ and apply the lowering operator to it to generate all of the different states \cite{condon_shortley,landau_lifschitz,cohen_tanoudji}. One works in  the coordinate representation and uses differential operators for $\hat L_-$. The fourth is to determine the rotation matrices as Wigner originally did~\cite{wigner,legendre2,gottfried_yau} and to take the $m'=0,m$ matrix element, which also leads to the spherical harmonics.
The final approach is to work with simple harmonic oscillator wavefunctions and the angular momentum representation discovered by Schwinger~\cite{schwinger}. This approach also allows one to determine the spherical harmonics from the rotation matrices~\cite{sakurai} in the same fashion as above.

Our approach, while closely related to methods four and five, is decidedly different. We focus on the meaning of the wavefunction as an overlap of the two eigenstates and develop the required rotation operators directly. The exponential disentangling operator identity is then naturally employed to get the final result. It does require a bit of operator algebra, but nothing that is too complicated. We feel it is a useful method six for deriving spherical harmonics.

Depending on the ordering of material in a course, this approach can fit in such a way that it could be the primary methodology for deriving spherical harmonics. In courses that begin with a discussion of the Stern-Gerlach experiment and spin, one would naturally discuss Pauli matrices and angular momentum early in the course. Once the angular momentum algebra and eigenstates are determined, then this approach fits well as a method to finish the spherical harmonics discussion.
The advantage is that it focuses directly on reinforcing operator methods and that it also 
helps use the algebra in a meaningful way to derive a nontrivial result.

We end this section with a discussion of some typical reinforcing exercises one can assign when
teaching this material. First, one can develop the angular momentum matrices for $l=1$ and show how the identity $\hat L_\alpha^3=\hat L_\alpha$ ($\alpha=x,$ $y$, or $z$) allows one to also compute the exponential of linear combinations of the angular momentum operators. Then, one can check the exponential disentangling operator identities for the faithful $l=1$ representation, which helps galvanize the
general group theory result. This exercise works in the following way. We begin with the standard $3\times 3$ representations for the angular momentum:
\begin{equation}
\hat L_x=\frac{\hbar}{\sqrt{2}}\left (\begin{array}{c c c} 0&1&0\\ 1&0&1\\ 0&1&0\end{array}\right ), \quad
\hat L_y=\frac{\hbar}{\sqrt{2}}\left (\begin{array}{c c c} 0&-\ri&0\\ \ri&0&-\ri\\ 0&\ri&0\end{array}\right ),\quad
{\rm and} \quad
\hat L_z=\hbar\left (\begin{array}{c c c} 1&0&0\\ 0&0&0\\ 0&0&-1\end{array}\right ).
\end{equation}
Then, using the fact that $\hat L_i^3=\hbar^2\hat L_i$, allows us to derive that
\begin{equation}
\re^{\ri \theta\frac{\hat L_y}{\hbar}}=\mathbb{I}+\ri\sin(\theta)\frac{\hat L_y}{\hbar}
+(\cos\theta-1)\left (\frac{\hat L_y}{\hbar}\right )^2=\left (\begin{array}{c c c}
\frac{1+\cos\theta}{2}&\frac{\sin\theta}{\sqrt{2}}&\frac{1-
\cos\theta}{2}\\
-\frac{\sin\theta}{\sqrt{2}}&\cos\theta&\frac{\sin\theta}{\sqrt{2}}\\
\frac{1-\cos\theta}{2}&-\frac{\sin\theta}{\sqrt{2}}&\frac{1+\cos\theta}{2}\end{array}\right ).
\end{equation}
Similarly, we have 
\begin{equation}
\hat L_+=\hbar\left ( \begin{array}{c c c} 0 & \sqrt{2} & 0\\
0 & 0 &\sqrt{2}\\
0 & 0 & 0\end{array}\right ) \quad {\rm and} \quad \hat L_-=\hbar\left ( \begin{array}{c c c}
0 & 0 & 0\\
\sqrt{2}&0&0\\
0&\sqrt{2}&0\end{array}\right )
\end{equation}
so that
\begin{equation}
\re^{-\tan\left (\frac{\theta}{2}\right )\frac{\hat L_-}{\hbar}}=\left ( \begin{array}{c c c}
1&0&0\\
-\sqrt{2}\tan\left (\frac{\theta}{2}\right )&1&0\\
\tan^2\left (\frac{\theta}{2}\right )&-\sqrt{2}\tan\left (\frac{\theta}{2}\right ) & 1\end{array}\right ),
\end{equation}
\begin{equation}
\re^{\ln\left [\cos^2\left (\frac{\theta}{2}\right)\right ]\frac{\hat L_z}{\hbar}}=\left ( \begin{array}{c c c}
\cos^2\left (\frac{\theta}{2}\right )&0&0\\
0&1&0\\
0&0&\sec^2\left (\frac{\theta}{2}\right )\end{array}\right ),
\end{equation}
and
\begin{equation}
\re^{\tan\left (\frac{\theta}{2}\right )\frac{\hat L_+}{\hbar}}=\left ( \begin{array}{c c c}
1&\sqrt{2}\tan\left (\frac{\theta}{2}\right )&\tan^2\left (\frac{\theta}{2}\right )\\
0&1&\sqrt{2}\tan\left (\frac{\theta}{2}\right )\\
0&0 & 1\end{array}\right ).
\end{equation}
Putting this together in the first form of the exponential disentangling operator identity, then gives
\begin{equation}
\re^{-\tan\left (\frac{\theta}{2}\right )\frac{\hat L_-}{\hbar}}
\re^{\ln\left [\cos^2\left (\frac{\theta}{2}\right)\right ]\frac{\hat L_z}{\hbar}}
\re^{\tan\left (\frac{\theta}{2}\right )\frac{\hat L_+}{\hbar}}=\left ( \begin{array}{c c c}
\frac{1+\cos\theta}{2}&\frac{1}{\sqrt{2}}\sin\theta&\frac{1-\cos\theta}{2}\\
-\frac{1}{\sqrt{2}}\sin\theta&\cos\theta&\frac{1}{\sqrt{2}}\sin\theta\\
\frac{1-\cos\theta}{2}&-\frac{1}{\sqrt{2}}\sin\theta&\frac{1+\cos\theta}{2}
\end{array} \right ),
\end{equation}
after using the half-angle formulas to convert all trigonometric functions to functions of $\theta$. This verifies the exponential disentangling operator identity for the $l=1$ multiplet. One can similarly derive the second form of the identity directly as well.

Second, one can ask students to explicitly extract the low $l$ 
spherical harmonics from the two different forms for the identity. This approach works well for $l=1$ and $l=2$. Since this expansion is easy to perform, we do not provide further details here. Finally, one could discuss how parity arises by taking $\theta\to \piup-\theta$ and $\phi\to\phi+\piup$; determining how $Y_l^m(\theta,\phi)$ changes under the parity transformation; and concluding that the parity operator acting on $|l,m\rangle$ returns $(-1)^l$ times the state. Once again, details are straightforward and are omitted.

\section{Summary}

In this work, we show how one can derive spherical harmonics without using any derivatives and argue that this approach brings in a number of important pedagogical advantages to the conventional treatment of orbital angular momentum (in particular, by introducing the powerful exponential disentangling operator identity). Obviously, one can use this approach to directly calculate the rotation matrices, and this has been known since the 1970s~\cite{arecchi,disentangle_rotation}. But the connection to use these results for spherical harmonics does not appear to have been made by anyone else, yet. 

We strongly feel that much of the beauty of quantum mechanics lies in how one can derive so many fundamental results from the different operator identities. We now see that this approach extends to deriving spherical harmonics as well. In our view, the more one can obtain results via operator methods, the better (just ask students to compare learning the differential equation approach for the simple harmonic oscillator versus the raising and lowering operator approach). In addition, this methodology teaches many of the strategies we employ later in research problems and it allows students to become facile in working with operators right from the start.

We end this work with a final comment to Prof. Stasyuk. We hope that he agrees with us about the importance of introducing operator methods such as these early into the curriculum. We feel he will appreciate this approach given his rich history of work with different operator-based formalisms. It is our honour to include this work in a volume dedicated to his 80th birthday.

\section*{Acknowledgements}

This work was supported by the National Science Foundation under grant number 
PHY-1620555. In addition, J.K.F. was also supported by the McDevitt bequest at Georgetown
University.

\ukrainianpart 

 \title{Розрахунок сферичних гармонік без похідних} 
 \author{М. Вайцман\refaddr{label1}, Дж.К. Фрірікс\refaddr{label2}} 
 \addresses{ 
 \addr{label1} просп. Е. Рошелл 3515, Лас-Вегас, Невада 89121, США 
 \addr{label2} Фізичний факультет, Джорджтаунський університет, \\вул. 37 \& О 
 NW, Вашинґтон, округ Колумбія 20057, США 
 } 

 \makeukrtitle 

 \begin{abstract} 
     \tolerance=3000% 
     Спосіб отримання сферичних гармонік приводиться однаково майже в кожному 
 підручнику чи на кожному занятті з квантової механіки. Як виявляється, його 
 трудно вслідкувати, важко зрозуміти й складно відтворити більшості студентів. 
 В цій роботі нами показано як обчислити сферичні гармоніки природнішим 
 способом з допомогою операторів та дивовижної тотожності, відомої як тотожність 
 експоненційного розплутування операторів (знаної в квантовій оптиці, але мало 
 застосовної десь інакше). Цей новий підхід виникає природнім чином після 
 введення діракових позначень, встановлення алгебри оператора кутового моменту 
 та визначення дії операторів збільшення та зменшення кутового моменту на 
 спільний базис власних функцій (як для $\hat L^2$ так і $\hat L_z$). 
     \keywords кутовий момент, сферичні гармоніки, операторні методи 

 \end{abstract}

\end{document}